\documentstyle{elsart}

\begin{document}

%%%%%%%%%%%%%%%%%%%%%%%%%%%%%%%%%%%%%%%%%%%%%%%%%%%%%%%%%%%%
%%%%%%% MANUSCRIPT %%%%%%%%%%%%%%%%%%%%%%%%%%%%%%%%%%%%%%%%%
%%%%%%%%%%%%%%%%%%%%%%%%%%%%%%%%%%%%%%%%%%%%%%%%%%%%%%%%%%%%

\runauthor{Ahrensmeier, Baier, Dirks}
\begin{frontmatter}
\title{Resonant decay of parity odd bubbles in hot hadronic matter}
\author[bielefeld]{D. Ahrensmeier}
\author[bielefeld]{\underline{R. Baier}\thanksref{email}}
\author[bielefeld]{M. Dirks}

\address[bielefeld]{Fakult\"{a}t f\"{u}r Physik, Universit\"{a}t
Bielefeld, D-33501 Bielefeld, Germany}

\thanks[email]{Electronic address: baier@physik.uni-bielefeld.de}

\date{Received today}
\maketitle
\begin{abstract}
 We investigate the decay of metastable states
 with broken CP-symmetry which have recently been proposed 
 by Kharzeev, Pisarski and Tytgat to form in hot hadronic matter. 
 We consider the efficiency of the amplification of the $\eta'$-field via
 parametric resonance, taking the backreaction into account.
 For times of the order $t\approx 10\: fm$, we find a particle density of 
 about $0.7\:/fm^3$ 
 and a correlation length of $\xi_{max}\approx 2.5 fm$.
 The corresponding momentum spectra show a non-thermal behaviour.
\hspace*{1ex}
\end{abstract} 
\end{frontmatter}

%%%%%%%%%%%%%%%%%%%%%%%%%%%%%%%%%%%%%%%%%%%%%%
%%%% SECTION I %%%%%%%%%%%%%%%%%%%%%%%%%%%%%%%
%%%%%%%%%%%%%%%%%%%%%%%%%%%%%%%%%%%%%%%%%%%%%%
\setcounter{equation}{0}
\setcounter{section}{0}

%%%%%%%%%%%%%%%%%%%%%%
\section{Introduction}
%%%%%%%%%%%%%%%%%%%%%%
 Recently, Kharzeev, Pisarski and Tytgat \cite{kha9804,kha0002}
 presented the following idea, based on the nontrivial
 topological structure of the QCD-vacuum \cite{don9200}:
 Metasta\-ble states which act like regions with a non-vanishing QCD 
 vacuum angle $\theta$ may be excited
 when matter undergoes the deconfining phase transition, provided it is of
 second order. In these false vacua, parity (and also CP)
 is spontaneously broken, a quality that could
 lead to experimental signatures like parity odd correlations of the
 produced particles.

 In the present work, we investigate the production of $\eta'$-particles 
 during the decay of the CP-odd metastable states, concentrating on the 
 amplification of the low momentum modes by parametric resonance when the 
 zero mode rolls down from the false to the true vacuum.
 This mechanism plays an important role for
 particle production in inflationary cosmology 
\cite{tra9000,boy9303,kof9400,gre9808,sin9509,kai9707,shi9700,cor9804} 
and has also been 
 investigated for the formation of Disoriented Chiral Condensates
 \cite{raj:rev,boy9501,mro9500,kai9801}
 and for Axions \cite{kol9800}.
 Here we discuss the efficiency of the amplification 
 mechanism for the production of $\eta$ from the decaying metastable 
 bubble. As a main result we derive the corresponding momentum 
 spectrum showing non-thermal behaviour. The integrated density of particles
 produced is estimated to be $0.7 fm^{-3}$. 

%%%%%%%%%%%%%%%%%%%%%%%%%%%%%%%%%%%%%%%%%%%%%%%%%%%%%%%%%%%%%%%
\section{Metastable states in hot hadronic matter}
%%%%%%%%%%%%%%%%%%%%%%%%%%%%%%%%%%%%%%%%%%%%%%%%%%%%%%%%%%%%%%%
 The idea is based on the effective Lagrangian of the 
 Witten-DiVecchia-Veneziano model \cite{ven7900,wit7900,vec8100}
\begin{eqnarray}
 {\cal L}_{eff} &=& \frac{f^2_{\pi}}{4}\left[tr(\partial_{\mu}U
        \partial_{\mu}U^+)  + tr(MU+MU^+) \right. \nonumber \\ \label{lagr}
        & & \hspace*{5cm}\left. -\frac{a}{N_c}\left(\theta - 
          \frac{i}{2}tr(\ln U -\ln U^+)\right)^2\right],
\end{eqnarray}
 which describes the low-energy dynamics of the 
 pseudoscalar mesons in the large $N_c$-limit of QCD.
 In the following, we consider the "real world"-vacuum angle $\theta\equiv 0$.
 Formation and decay of a non-zero $\theta$-vacuum and its signatures
 are investigated in \cite{zhit}.
 The $N_f \times N_f$-matrix $U$ in Eq.~(\ref{lagr})
 describes the meson fields, and the
 term containing the mass matrix $M$ provides the explicit soft
 breaking of chiral symmetry due to the quark masses. $M$ can be written 
 as $M_{ij}= \mu^2_i \delta_{ij}$,
 where the diagonal elements are chosen as
 $\mu^2_{1}\equiv \mu^2_{2}\equiv m^2_{\pi}$ and 
 $\mu^2_{3}\equiv  2 m^2_K - m^2_{\pi}$.
 With the parametrization $U=\exp(i\phi/f_{\pi})$,
 the matrix $\phi$ representing the singlet and the octet meson fields reads 
\begin{equation}
 \phi = \sqrt{\frac{2}{3}}\eta_1 {\bf 1} +
 \left( \begin{array}{ccc}
  \pi^0 + \frac{\eta_8}{\sqrt{3}} & \sqrt{2}\pi^+ & \sqrt{2}K^+ \\
  \sqrt{2}\pi^- & -\pi^0 +\frac{\eta_8}{\sqrt{3}} & \sqrt{2}K^0 \\
  \sqrt{2}K^- & \sqrt{2}\bar{K^0} & -\frac{2}{\sqrt{3}}\eta_8
  \end{array} \right).
\end{equation}
 The last term in the effective Lagrangian  reflects the $U_A(1)$-anomaly 
 by giving a mass
 to the singlet, even in the chiral limit of vanishing 
 quark masses. The parameter $a=2 N_f \lambda_{YM}/f^2_{\pi}$
 represents the topological susceptibility.

 Finite temperature is introduced into the model
 via the temperature dependence of the parameters: For $T=0$, the value 
 $a=m^2_{\eta}+m^2_{\eta'}-2m^2_K \simeq 0.726\: GeV^2$
 can be determined from experimental results \cite{ven7900,wit7900,vec8100},
 and $\mu^2 \equiv (\mu^2_1 + \mu^2_2 + \mu^2_3)/3
       = (m^2_{\pi}+2m^2_K)/3 \simeq 0.171\: GeV^2$. The corresponding value
 of the pion decay constant is $f_{\pi}\simeq 93\: MeV$.
 
 When the temperature approaches the temperature $T_d$ of the phase transition,
 $a(T)$ goes to zero, indicating the effective restoration of 
 the $U_A(1)$-symmetry for $T>T_d$ with enhanced $\eta$ - and $\eta'$ - 
 production as a possible signature \cite{kap:eta,huang:eta}.
 We use the temperature of the deconfinement phase transition,
 $T_d$, following the assumption of \cite{kha9804} that any other phase
 transition of this model also occurs at $T_d$.
 According to mean field estimates \cite{kha9804}, 
 $ a(T)\propto (T_d - T) $ and $\mu^2(T)\propto (T_d-T)^{-1/2}$ near the
 phase transition.

 In order to get an analytically tractable approximation that allows 
 some insight into the physical processes, we consider only the singlet, 
 which is the main component of the $\eta'$-meson.
 Using $N_c=3$, the effective Lagrangian simplifies to 
\begin{equation}\label{ourL}
 {\cal L}_{s}=\frac{1}{2}(\partial_{\mu}\eta_1)(\partial^{\mu}\eta_1) 
 +\frac{3}{2}f^2_{\pi}\mu^2\cos\left(\sqrt{\frac{2}{3}}\frac{\eta_1}{f_{\pi}}\right)
 -\frac{a}{2}\eta^2_1.
\end{equation}
 The singlet effective potential
\begin{equation}\label{Veff}
 V\left(\frac{\eta}{f}\right)\equiv
 \frac{V_s}{f^2\mu^2}\left(\frac{\eta}{f}\right) = 
 - \cos\frac{\eta}{f} 
 +\frac{a}{2\mu^2}\frac{\eta^2}{f^2},
\end{equation}
 where we have dropped the index $1$ in the notation of the singlet
 and introduced $f\equiv \sqrt{3/2}f_{\pi}$ for convenience, can be  considered
 as a projection of the full effective potential.
 It takes  three qualitatively 
 different shapes, depending on the temperature, more
 specifically, on the value of $a(T)/\mu^2(T)$. For {\em large} values of 
 $a/\mu^2$, corresponding to {\em small} $T$, the potential
 is a more or less deformed parabola (see Fig.~\ref{fig:pot}a). 
 For $a/\mu^2=0$, i.e.
 $T>T_d$, the potential is periodic with minima at $\eta/f = 0,2\pi,4\pi...$,
 which are all equivalent  true vacua (see Fig.~\ref{fig:pot}d).
 And for a {\em small, nonzero} value of $a/\mu^2$, corresponding to a temperature 
 just below the phase transition, metastable states appear in the potential 
 which are distinct from the true vacuum (see Fig.~\ref{fig:pot}c).
 One recognizes that the false vacua are odd 
 under parity transformation, but even under charge 
 conjugation, and therefore odd under CP \cite{kha9804}.
 
 One can show that the effective potential exhibits metastable states also
 in the case of nonvanishing $\pi-,\eta_8-\:\mbox{and}\:K-$fields, 
 provided that
 $m_{\pi}\neq 0$ \cite{kha0001}. But only the singlet is responsible for the
 formation of these  states, in the sense that it is the only field
 which appears in the term of the effective potential which includes the
 topological susceptibility. 
%%%%%%%%%%%%%%%%%%%%%%%%%%%%%%%%%%%%%%%%%%%%
\section{The decay of the metastable states}
%%%%%%%%%%%%%%%%%%%%%%%%%%%%%%%%%%%%%%%%%%%%

%%%%%%%%%%%%%%%%%%%%%%%%%%%%%%%%%%
\subsection{Dynamics of the model}
%%%%%%%%%%%%%%%%%%%%%%%%%%%%%%%%%%
 Applying the model to the description of a heavy ion
 collision, we suppose that during the phase transition, the potential
 changes its shape from the parabola (Fig.~\ref{fig:pot}a), corresponding to the low
 temperature phase, to that of the high temperature phase (Fig.~\ref{fig:pot}d), 
 and back to the low temperature phase, but this time including
 metastable states that have been created at a temperature just below the phase
 transition.
 After the phase transition, but still at high temperature, the
 $\eta'-$field may be trapped in a metastable state, forming a "CP-odd
 bubble inside the hadronic phase" in the language of \cite{kha9804}.
 Considering the  effective potential as a function of temperature,
 one recognizes that there exists a value $a/\mu^2(T_{sp})$ for which the
 local minimum turns into a saddle point (see Fig.~\ref{fig:pot}b). 
For the lowest false
 vacuum, this happens for $a/\mu^2(T_{sp}) = 0.217$ at the value
 $(\eta/f)_{sp}=4.493$.
 Assuming the proportionality $a(T)/\mu^2(T)\propto (T_d -T)^{3/2}$ to hold
 even far from $T_d$, the temperature corresponding to this special case of 
 the potential can be estimated to be $T_{sp} \approx 0.86\: T_d$.
 Once the saddle point temperature is reached and the barrier has disappeared,
 the $\eta'-$field starts rolling down 
 into the true vacuum and oscillating around it, while 
 energy is transferred from the mean value of the field into its
 quantum fluctuations.
 The resulting growth of the occupation numbers of the quantum fluctuations is
 interpreted as particle production with respect to the true, CP even, vacuum,
 whereas the zero mode is damped. This process is investigated in more detail
 in the following.
 
 A crucial assumption for this picture is that
 at least after the saddle point temperature has been
 reached, the effective potential changes slowly compared to the motion of
 the field, so that the oscillations can be considered as taking place in a 
 static potential, frozen at $T=T_{sp}$. Accordingly, the ratio $a/\mu^2 =
 0.217$ is kept fixed during the evolution in time, and the value of $f$ as well.
 The solutions we find (see below) show
 that the zero mode reaches the true vacuum for the first time 
 after  $1-2\:fm$, whereas the relevant time scale to be considered for the
 particle production is of the order of $10\:fm$.

%%%%%%%%%%%%%%%%%%%%%%%%%%%%%%%%
\subsection{Evolution equations}
%%%%%%%%%%%%%%%%%%%%%%%%%%%%%%%%
 To study the dynamics of the decay process, we decompose the field into its 
 expectation value $\varphi(t)\equiv \langle \eta(\vec{x},t)\rangle$ and
 fluctuations $\chi$ about it,
\begin{equation}
 \eta(\vec{x},t)=\varphi(t)+\chi(\vec{x},t)
\end{equation}
 with $\langle \chi(\vec{x},t)\rangle = 0$,
 so that the "physical" fluctuations are given by the usual expression
 $\langle(\chi -\langle\chi\rangle)^2\rangle = \langle\chi^2\rangle$. 
 With this decomposition, following the standard method as described in, e.g.,
 \cite{boy9303,kof9400,sin9509,cor9804}, we obtain two 
 evolution equations from the effective 
 Lagrangian (\ref{ourL}). The effect of the quantum 
 fluctuations at lowest order is taken into account with a Hartree-type
 approximation \cite{boy9303,sin9509,cor9804} that consists of the 
 factorization
 $\chi^3\rightarrow 3\langle\chi^2\rangle\chi$ and
 $\chi^2\rightarrow \langle\chi^2\rangle$ and the self-consistency condition
 represented by Eq.~(\ref{expval}) given below; terms of higher order
 in $\chi$ are neglected. Taking the expectation value $\langle\rangle$,
 we arrive at the equation for the zero mode
\begin{equation}\label{zero}
 \frac{d^2}{d\tau^2}\frac{\varphi(\tau)}{f}+
  \left(1-\frac{\langle\chi^2\rangle_{\tau}}{2f^2}\right)\sin\frac{\varphi(\tau)}{f}
  +\frac{a}{\mu^2}\frac{\varphi(\tau)}{f}=0,
\end{equation}
 written for the dimensionless mean field $\varphi/f$, where the 
 dimensionless time variable $\tau\equiv\mu t$ is introduced.
 The expectation value of the fluctuations
\begin{equation}\label{expval}
 \langle\chi^2\rangle_\tau\equiv\int\frac{d^3 k}{(2\pi)^3}|\chi_k(\tau)|^2 
 \end{equation}
 is calculated from the mode functions $\chi_k (\tau)$ which are known from the
 Fourier expansion of the field.
 The integration is done over the interval of momenta for which
 the quantum fluctuations are enhanced.

 We use the definition (\ref{expval}) for the back reaction, although it
 does not fulfill $\langle\chi^2\rangle_0=0$, because it reflects the
 background of the calculation, the closed-time-path technique which is usually
 applied for the description of out-of-equilibrium dynamics in field theory.  
 In this framework, the
 fluctuations are defined  as the two-point correlation functions or 
 equal-time Green's functions \cite{boy9303}
\begin{equation}
 \langle\chi^2\rangle_\tau\propto\int\frac{d^3k}{(2\pi)^3}G_k(\tau,\tau).
\end{equation}

 The mode functions satisfy the mode equation
\begin{equation}\label{modeeq}
 \frac{d^2}{d\tau^2}\chi_{\kappa}(\tau)
 +\omega^2_{\kappa}(\tau)\chi_{\kappa}(\tau)=0,
\end{equation}
 written for  the dimensionless mode functions 
 $\chi_{\kappa}\equiv\sqrt{\mu}\chi_k$, where 
\begin{equation}\label{omega}
 \omega^2_{\kappa}(\tau)=\kappa^2
 +\left(1-\frac{\langle\chi^2\rangle_{\tau}}{2f^2}\right)\cos\frac{\varphi(\tau)}{f}
 +\frac{a}{\mu^2}
\end{equation}
 is the time-dependent frequency squared with $\kappa\equiv k/\mu$.
 They are used for the calculation of the
 particle production:  The induced particle number density
\begin{equation}\label{num}
 n(\tau)=\int\frac{d^3 k}{(2\pi)^3}n_{k}(\tau)
\end{equation}
is obtained from the spectral particle number density 
\begin{equation}\label{numk}
 n_{\kappa}(\tau)=\frac{\omega_{\kappa}}{2}\left(|\chi_{\kappa}(\tau)|^2 +
          \frac{|\dot{\chi}_{\kappa}(\tau)|^2}{\omega_{\kappa}^2}\right) - 
                  \frac{1}{2},
\end{equation}
 which can be calculated from the mode functions. Since the unstable solutions
 of the mode equations contribute to the resonant particle production,
 the integration in (\ref{num}) is carried out over the momenta corresponding
 to these amplified  solutions, as in Eq.~(\ref{expval}).
 For practical purposes, we use $\bar{\omega}^2_{\kappa}\equiv \kappa^2 + a/\mu^2$ in 
 definition (\ref{numk}), 
 because it keeps the particle number positive definite
 for all values of $\tau$ and minimizes the fluctuations
 in the particle number. Accordingly, the initial values of the mode
 functions are chosen as 
 $\chi_{\kappa}(0)=1/\sqrt{2\bar{\omega}_{\kappa}}$ and
 $\dot{\chi}_{\kappa}(0)=-i\sqrt{\bar{\omega}_{\kappa}/2}$
 %has to be modified then, using $\bar{\omega}_{\kappa}$ too,
 such that $n_{\kappa}(0)=0$. However, the results depend only weakly on
 the choice of $\omega_{\kappa}$ both in the initial conditions and in the
 definition of the particle number. 
 The choice for the initial conditions of the zero mode is suggested by the
 model: $\phi$  is "released" right below the saddle point 
 $\varphi_{sp}/f=4.493$ with zero velocity, i.e. $\dot{\varphi}(0)=0$. 
 For the numerical results shown below, we used the starting value
 $\varphi(0)/f = 4.2$.

%%%%%%%%%%%%%%%%%%%%%%%%%%%%%%%%%%
\subsection{Solutions and results}
%%%%%%%%%%%%%%%%%%%%%%%%%%%%%%%%%%

 Considering the coupled system (\ref{zero}) and (\ref{modeeq}) 
 with (\ref{omega}), one recognizes
 that the expectation value transfers energy  to the mode functions via a time
 dependent frequency, and the mode functions in turn modify the equation of
 motion for the zero mode. A natural first step in the solution of this system
 is to neglect the fluctuation terms in  both equations,
 obtaining the "classical" evolution equations with 
 $\langle\chi^2\rangle_{\tau}=0$   
 where  the mode functions do not react back on the zero mode. If the variable
 frequency
\begin{equation}
 \omega_{\kappa,cl}^2=\kappa^2 + \cos\frac{\varphi(\tau)}{f} + \frac{a}{\mu^2}
\end{equation}
 in the mode equation depends {\em periodically} on time, 
 according to Floquet's theorem \cite{inc5600} it has quasiperiodic solutions of the form
\begin{equation}\label{Floq}
 \chi^{cl}_{\kappa}(\tau)=\exp(\mu_{\kappa} \tau)P(\tau), 
\end{equation}
 where the characteristic exponent $\mu_{\kappa}$ depends on
 $\omega_{\kappa,cl}$, and $P(\tau)$ is a periodic function of the same period as 
 $\omega_{\kappa,cl}$, usually normalized to have unit amplitude. For certain 
 values of the parameter $\omega_{\kappa,cl}$, there exist so called resonance 
 bands that contain the exponentially
 growing (or resonant) solutions with a real characteristic exponent
 $\mu_{\kappa}$. These resonance bands are known, for example, for the solutions
 of the Mathieu- or the Lam\'{e}-equation.

 An analytic solution of the classical system can be found with a
 Sine-Gordon-equation \cite{gre9808}, which consists essentially of setting
 $a/\mu^2\approx 0$ in our case. The corresponding approximate zero mode
 equation is given by
\begin{equation}
 \frac{\ddot{\varphi}_{cl}(\tau)}{f}+\rho\sin\frac{\varphi_{cl}(\tau)}{\rho f}=0,
\end{equation}
 where $\rho=1.43$ represents the radius of the potential and is needed to
 adjust it to the original potential. It has the solution
\begin{equation}
 \frac{\varphi_{cl}(\tau)}{f}=2\rho\arctan
 \left(\sqrt{\frac{\epsilon}{2-\epsilon}}cn(\tau,m)\right)
\end{equation}
 with $\epsilon\equiv 1-\cos(\varphi(0)/\rho f)$,  and
 $m=\epsilon/2$ is the square of the modulus of the Jacobian elliptic 
 function $cn$. The resulting mode equation is a Lam\'{e}-equation
 with the resonance band \cite{gre9808} $0\leq\kappa^2\leq\epsilon/2$.
 The resulting $\mu_{\kappa}$ is shown in Fig.~\ref{fig:mu} in comparison with the 
 result of the numerical calculation of $\mu_{\kappa}$ for the solution
 of the classical mode equation.
 In the range of the single resonance band of the Lam\'{e}-equation,
 $0\leq\kappa^2\leq0.97$, the numerical calculation yields 
 {\em three} resonance bands, and at least two for
 $\kappa^2 > 1$.
 (This structure reminds of the Mathieu-equation
 which has more than one resonance band.) The maximum value
 $\mu_{\kappa,max}\simeq 0.42$ is found around
 $\kappa^2\approx0.1$, and the maximum value for the
 Sine-Gordon-solution, $\mu_{\kappa,max}^{sg}\simeq 0.5$, yields an
 acceptable approximation .

 The momentum spectrum of the produced particles  reflects the structure 
 of the resonance bands. When the general expression for the mode functions
 (\ref{Floq}) is inserted into Eq.~(\ref{numk}), it is obvious that the particle
 production is  essentially determined by $\mu_{\kappa}$:
\begin{equation}\label{lnnk}
 \ln n_{\kappa}(\tau)-\ln\frac{\omega_{\kappa}(\tau)}{2}\simeq 2\mu_{\kappa}\tau.
\end{equation}

 In the classical approximation discussed above, the back reaction of 
 the fluctuations on the zero mode (and also on the fluctuations themselves)
 is ignored. This implies that the approximation is not energy conserving,
 which can immediately be seen from the zero mode solution which is not 
 damped although it is supposed to transfer its energy to the mode functions, 
 which indeed grow exponentially.
 Since the mode functions are not decoupled from the zero mode, at least the 
 other direction of energy transfer is reflected by the approximation:
 The larger the initial value $\varphi(0)$, the more energy can be transferred,
 and the larger the mode functions and the particle number grow.

 The evolution equations including the back reaction are solved 
 numerically. For the evaluation, we use $f(T_{sp})\simeq f(0)$ and
 $\mu(T_{sp})\simeq 676\:MeV$. Because the value of $\mu$ in our approximation
 is only fixed by the value of $T_{sp}$, which depends on our restriction
 to the singlet case and may not correspond to the real physical situation,
 we also try $\mu(T=0)=412\:MeV$ as the other limit of the range of possible
 values for $\mu(T)$.

 Although the inclusion of the back reaction destroys the parametric resonance,
 it does not suppress the particle production completely.
 The particle number (Fig.~\ref{fig:ntau}) reaches its asymptotic value
 for $t_{end}\approx 10\:fm$. 
 At about the same time, the value of the back reaction term 
 $\langle\chi^2\rangle_\tau/2f^2$
 grows larger than 1,
  indicating that the approximations made become inappropriate for the
 description of the dynamics for larger times (cf. Eq.~(\ref{zero})).
 One recognizes from 
Fig.~\ref{fig:ntau} that
 the choice of $\mu(T)$ has some influence on $t_{end}$; the asymptotic
 number density of the produced particles lies between 
 $0.7 fm^{-3}$ and $0.9 fm^{-3}$.
 The absolute number of particles can be calculated from this density
 if the size of the CP-odd bubbles is known. For example, for a domain with 
 radius $5\:fm$, as suggested in  \cite{kha9906}, we estimate a yield of
 90-100 particles from our model.

 A measure for the size of correlated domains of particles is the correlation
 length $\xi$, which can be calculated from the two-point correlation function
\begin{equation}
 D(x,\tau)\propto\int\frac{d\kappa\:\kappa^2}{2\pi^2}j_0(\kappa x)
 |\chi_{\kappa}(\tau)|^2,
\end{equation}
 where $x=\mu r$ denotes the dimensionless length. For large $x$,
\begin{equation}
 D(x,\tau)\propto\exp\left(\frac{-x^2}{\xi^2(\tau)}\right),
\end{equation}
 from which we estimate the correlation length for $t_{end}$. As a lower limit
 for the domain radius, we obtain
 $\xi\simeq 2.4 \:fm$ for $\mu(T=0)$ and $\xi\simeq 1.3 \: fm$ for
 $\mu(T_{sp})$.

 The momentum spectrum of the produced particles at $t_{end}\simeq 10\:fm$
 is shown in Fig.~\ref{fig:spec}. Its maximum is located at about the  value of
 $\kappa$ as the maximum of $\mu_{\kappa}$ in Fig.~\ref{fig:mu}, which
 had been calculated
 for the system {\em without} back reaction. The spectrum deviates 
 significantly from a thermal Bose-Einstein distribution.

\begin{ack}
 R.B. thanks the E. Schr\"odinger Institute, Vienna, for support 
 during the workshop "Quantization, Generalized BRS Cohomologies
 and Anomalies", where this work has been encouraged from discussions
 with M.H.G. Tytgat. We thank especially D. Kharzeev for useful 
 remarks and suggestions. This research is supported in part by DFG, contract
 Ka 1198/4-1.\\[1cm]
\end{ack}

\clearpage\newpage
\section*{Figure Captions}

\begin{figure}[h]
\caption{\label{fig:pot} Form of the  singlet effective potential 
for different temperatures: (a) $T = 0$, (b) $T=T_{sp}$, 
(c) $T_{sp} \le T \le T_d$, (d) $T_d < T$ .  }
\end{figure}

\begin{figure}[h]
\caption{\label{fig:mu} The characteristic exponent $\mu_{\kappa}$ for the solutions of the 
 classical evolution equations (solid line) and for the analytical
 solution of the Sine-Gordon-approximation (dotted line).
}
\end{figure}

\begin{figure}[h]
\caption{\label{fig:ntau} Particle number density per $fm^3$ as a 
 function of time for
 $\mu(T=0)=412\:MeV$ (dashed line) and $\mu(T_{sp})\simeq 676\:MeV$
 (full line).}
\end{figure}

\begin{figure}[h]
\caption{\label{fig:spec} Momentum spectrum of produced particles at $t_{end}\simeq 10\:fm$, 
 including back reaction and using $\mu(T=0)=412\:MeV$.}
\end{figure}

\end{document}